\documentclass{ws-procs9x6}

\newcommand{\be} {\begin{eqnarray*}}
\newcommand{\ee} {\end{eqnarray*}}

\newcommand{\bcen}{\begin{center}}
\newcommand{\ecen}{\end{center}}
\newcommand{\beq}{\begin{equation}}
\newcommand{\eeq}{\end{equation}}
\newcommand{\bea}{\begin{eqnarray}}
\newcommand{\eea}{\end{eqnarray}}
\newcommand{\ba}{\begin{array}}
\newcommand{\ea}{\end{array}}
\newcommand{\bann}{\begin{eqnarray*}}
\newcommand{\eann}{\end{eqnarray*}}

\begin{document}

\title{Cooling of Neutron stars with color superconducting quark cores
\footnote{\uppercase{C}onference \uppercase{P}roceedings of the
\uppercase{KIAS-APCTP} \uppercase{I}nternational \uppercase{S}ymposium in
\uppercase{A}stro-\uppercase{H}adron \uppercase{P}hysics
\uppercase{C}ompact \uppercase{S}tars: \uppercase{Q}uest for
\uppercase{N}ew \uppercase{S}tates of \uppercase{D}ense \uppercase{M}atter}}

\author{ D.~Blaschke}
\address{Fachbereich Physik, Universit\"at Rostock,
        D--18051 Rostock, Germany\\
        Bogoliubov Laboratory for Theoretical Physics, JINR Dubna,
        141980 Dubna, Russia\\
        email: david@thsun1.jinr.ru}
\author{ D. N. ~Voskresensky}
\address{
         Theory Division, GSI mbH,
         D--64291 Darmstadt, Germany\\
         Moscow Institute for Physics and Engineering,
         115409 Moscow, Russia}
\author{ H.~Grigorian}
\address{FB Physik, Universit\"at Rostock,
        D--18051 Rostock, Germany\\
        Department of Physics, Yerevan State University, Alex
        Manoogian Str. 1, 375025 Yerevan, Armenia
        }
\maketitle

\abstracts{
We show that within a recently developed nonlocal, chiral quark model
the critical densities for a phase transition to color superconducting
quark matter under neutron star conditions can be low enough that these
phases occur in compact star configurations with masses below $1.4~M_\odot$.
We study the cooling of these objects in isolation for different values
of the gravitational mass and thus different composition and structure of
the interior.
Our equation of state allows for a  2SC phase with a large quark
gap $\Delta \sim 100~$MeV for $u$ and $d$ quarks of two
colors, a normal quark matter phase and their coexistence in a mixed phase
within the hybrid star interior.  We argue that, if  the phases
with unpaired quarks were allowed, the corresponding hybrid stars
would cool too fast to describe the neutron star cooling data
existing by today. We incorporate other attractive channels
permitting a weak pairing of the residual quarks which remained unpaired
in the 2SC phase and demonstrate that the model does not
contradict the cooling data if the weak pairing gaps are of the order
of $0.1~$ MeV.
}

\section{Introduction}
In the recent paper \cite{bgv2004}, hereafter BGV, we have reinvestigated
the cooling of neutron stars (NS) within a purely hadron model, i.e.,
ignoring the possibility of quark cores in NS interiors.
We have demonstrated
that the NS cooling data available by today can be well explained
within the {\em "Nuclear medium cooling scenario"}, i.e., if one
includes medium effects in the emissivity and takes into account
a suppression of  the $3P_2$ neutron gap.
In a subsequent work \cite{bgv2004q} we have shown that this result does
not exclude the possibility that neutron stars might possess large
quark matter cores that extend up to more than half of the star radius.
Such a hybrid structure gives room for a whole variety of additional
scenarios of compact star cooling which fall into two classes:
either nuclear and quark matter phases have similar cooling behavior
(homgeneous cooling) or the faster cooling of the one phase is compensated
by the slower cooling of the other (inhomogeneous cooling).
In the present contribution we will report on our results within the former,
homgeneous cooling scenario of hybrid stars and what implications the
comparison with present-day cooling data may provide for the EoS and
transport properties of quark matter.

\section{Color superconductivity}

The quark-quark interaction in the color anti-triplet channel is
attractive driving the pairing with a large zero-temperature
pairing gap $\Delta\sim 100$~MeV for the quark chemical
potential $\mu_q \sim (300\div 500)$~MeV, cf.
\cite{arw98,r+98}, see review \cite{RW} and Refs therein.
The attraction comes either from the one-gluon exchange, or from a
non-perturbative 4-point interaction motivated by instantons
\cite{dfl} or from non-perturbative gluon propagators \cite{br}.
Various phases are possible. The so called 2-flavor color
superconductivity (2SC) phase allows for unpaired quarks of one
color, say blue. There may also exist a color-flavor locked
(CFL) phase \cite{arw99} for not too large values of the dymanical
strange quark mass or other words for very large values of the baryon
chemical potential \cite{abr99}, where the color superconductivity
(CSC) is complete in the sense that the diquark condensation
produces a gap for quarks of all three colors and flavors. The
value of the gap is of the same order of magnitude as that in the
two-flavor case. There exist other attractive quark pairing
channels for quarks that can't participate in
2SC and CFL pairing.
These weak pairing channels are characterized by gaps
typically in the interval $\sim 10~$ keV $\div 1~$MeV,
and a prominent example which will be used in the present work is
the spin-1 pairing channel of single color diquarks in the isospin
singlet state, also named as color-spin locking (CSL) phase.

The high-density phases of QCD at low temperatures may exist in
the interiors of hybrid stars affecting their cooling, rotation and
magnetic field evolution, cf. \cite{bkv,ppls,bgv,bss,IB}.

\section{Hybrid stars}

In describing the hadronic part of the hybrid star we  exploit
the Argonne $V18+\delta v+UIX^*$ model of the EoS given in \cite{APR98}, which
is based on the most recent models for the nucleon-nucleon interaction with the
inclusion of a parameterized three-body force and relativistic boost corrections.
Actually we adopt here an analytic parameterization of this model by Heiselberg
and Hjorth-Jensen \cite{HJ99}, hereafter HHJ.
The latter uses a compressional part with the incompressibility $K\simeq 240$~MeV,
and a symmetry energy fitted to the data around nuclear saturation density that smoothly
incorporates causality at high densities. The density dependence
of the symmetry energy is very important since it determines the
value of the threshold density for the DU process ($n_c^{\rm DU}$).
The HHJ EoS fits the symmetry energy  to the original
Argonne $V18+\delta v +UIX^*$ model yielding $n_c^{\rm
DU}\simeq~5.19~n_0$ ($M_c^{\rm DU}\simeq 1.839~M_{\odot}$).

The 2SC phase occurs at lower baryon densities than the CFL phase,
see \cite{SRP,NBO}. For applications to compact
stars the omission of the strange quark flavor is justified by the
fact that central chemical potentials in star configurations do
barely reach the threshold value at which the mass gap for strange
quarks breaks down and they appear in the system \cite{GBKG}.

We will focus on the model of the quark EoS developed in
\cite{BFGO}. The Gaussian, Lorentzian and  NJL type cutoff
formfactors were studied. The  Lorentzian interpolates between a
soft (Gaussian type, $\alpha \sim 2$), and a hard (NJL, $\alpha
>30$) depending on the value of the parameter $\alpha$.
We will further work out two possibilities of the Gaussian and the
Lorentzian formfactors.

\begin{figure}[ht]
\vspace{-0.5cm}
\centerline{
\psfig{figure=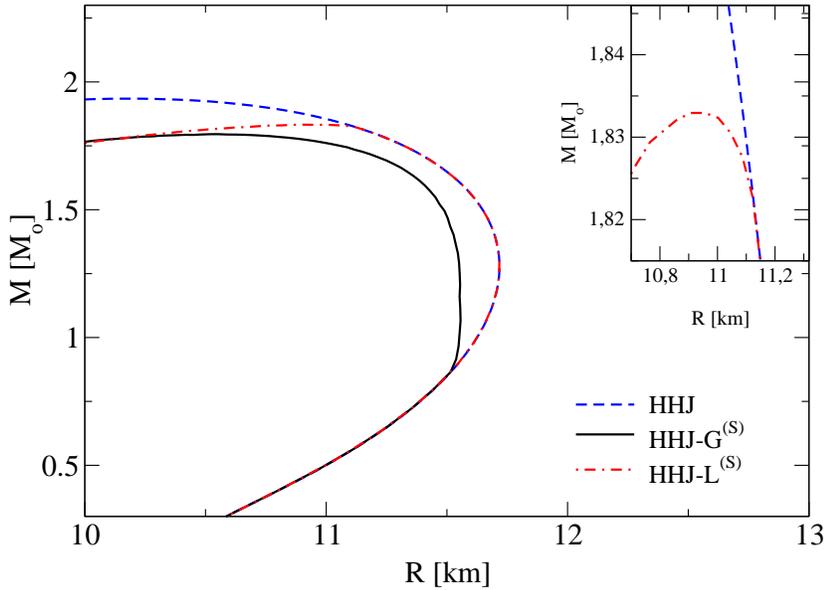,height=0.85\textwidth,angle=-90}}
\caption{Mass - radius 
relations for compact star configurations with different EoS:
purely hadronic star with HHJ EoS (dashed line),
stable hybrid stars with HHJ - SM$_{\rm G}^{\rm (S)}$ EoS (solid line) and
with HHJ - SM$_{\rm L}^{\rm (S)}$ EoS (dash-dotted line).
\label{fig:stab}}
\end{figure}

In some density interval at the first order phase transition there
may appear the region of the mixed phase, see \cite{G92}. Ref.
\cite{G92} disregarded finite size effects, such as surface
tension and charge screening. Refs \cite{VYT02} on the example of
the hadron-quark mixed phase have demonstrated that finite size
effects might play a crucial role substantially narrowing the
region of the mixed phase or even forbidding its appearance.
Therefore we omit the possibility of the hadron-quark mixed phase in our
model where the quark phase arises by the Maxwell construction.
For the case of the Gaussian formfactor the quark core appears for
$n>n_c=0.34 {\rm fm}^{-3}$ ($M>0.852~M_\odot$), for the Lorentzian formfactor 
$n>n_c=0.80 {\rm fm}^{-3}$ ($M>1.81~M_\odot$).
In the following, we will not further discuss the Lorentzian case which 
gives a marginal quark core in a small mass range only, see
Fig. \ref{fig:stab}.

A large difference between chemical potentials of $u$ and $d$
quarks forbids the pure 2SC phase, cf. \cite{BFGO}. 
The CFL phase is still not permitted at such
densities. Ignoring the possibility of a weak coupling  we have  two
possibilities: either the quark matter for $n>n_c$ is in the normal
phase, or there appears a region of the 2SC -- normal quark mixed
phase. Following \cite{G92}, Ref. \cite{GBKG} considered the
latter possibility disregarding finite size effects and has found
the possibility of a wide region (for Gaussian formfactor) of the
2SC -- normal quark mixed phase instead of a pure 2SC phase. In the
given case arguments of \cite{VYT02} are relaxed since the surface
tension on the 2SC -- normal quark boundary should be much less
compared to that for the quark -- hadron boundary. Indeed, the
surface tension is proportional to the difference of the energies
of two phases, being very small in the given case, $\propto(\Delta
/\mu_q )^2 \ll 1$.

In Fig. \ref{fig:stab} we present the mass-radius relation for hybrid stars
with HHJ vs. SM EoS. Two sets of configurations given by Gaussian 
(solid lines) and Lorentzian (dash-dotted lines) formfactors are stable, 
see also Ref. \cite{BFGO},
where similar results for the nonlinear Walecka model EoS (RMF) have been 
found.


If one switches on the possibility of the weak coupling, e.g. the
CSL pairing channel, see \cite{ABCC,NBO,BHO} all the quarks in
the normal phase may acquire a corresponding spin-1 pairing gap,
typically $\Delta \sim 10$~keV $\div 10$~MeV. 
In such a way all the quarks may get paired, either strongly in the 
2SC channel or weakly in the CSL one.

\section{Cooling}
For the calculation of the cooling of the hadron part of the
hybrid star we adopt the same model as in BGV.
The main processes are medium modified Urca (MMU) and pair
breaking and formation (PBF) processes. The HHJ EoS was adopted.
In Fig. 2 (Fig. 20 of BGV) we show the cooling of
different NS calculated within the hadron model of BGV.
We use a fit-law for the relation between surface and interior
temperatures, see BGV. Possibilities of the pion
condensation and of the other so called exotic processes are for
simplicity suppressed. Direct Urca is irrelevant in this model up
to very large NS mass $M>1.839~M_{\odot}$. $1S_0$ neutron and
proton gaps are taken the same as in paper \cite{TT04}, whereas
$3P_2$ neutron gap is suppressed by the factor $0.1$, see Fig. 4
of BGV.

For the calculation of the cooling of the quark part of the hybrid
star we are basing on the model \cite{bgv}. We include the most
efficient processes: the quark direct Urca (QDU) processes on
unpaired quarks, the quark modified Urca (QMU), the quark
bremsstrahlung (QB), the electron bremsstrahlung (EB) Following
\cite{JP02} we include the emissivity of the quark pair formation
and breaking (QPFB) processes. The specific heat incorporates the
quark contribution, the electron contribution and the gluon-photon
contribution. 
In the CSL phase \cite{BHO} the specific heat is proportional to $T^2$.
This new term does not significantly affect the total specific heat 
since the main contribution comes from electrons.
The heat conductivity contains quark, electron and gluon terms.

\begin{figure}[ht]
\vspace{-0.5cm}
\centerline{
\psfig{figure=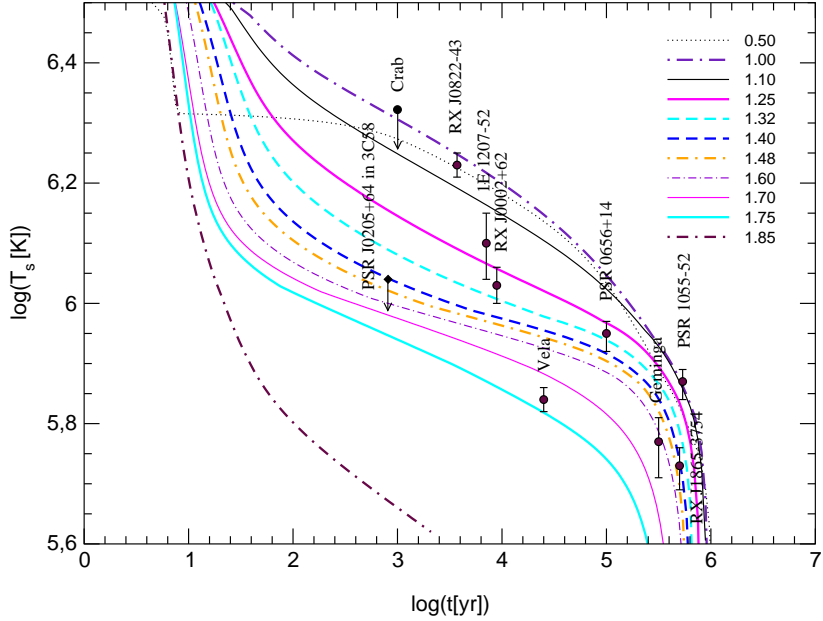,height=0.9\textwidth,angle=-90}}
\caption{Cooling curves according to the nuclear medium cooling scenario,
see Fig. 20 of BGV.
The labels correspond to the gravitational masses of the configurations in 
units of the solar mass.}
\label{fig:cool-h}
\end{figure}

We are basing on the picture presented in Fig. \ref{fig:cool-h} and add the
contribution of the quark core. For the Lorentzian formfactor the
quark core appears only for $M>1.81~M_\odot$, see Fig. \ref{fig:stab}. 
The existing cooling data are not affected, thereby.

For the Gaussian formfactor the quark core occurs already for
$M>0.852~M_\odot$ according to the model \cite{BFGO}, see Fig. \ref{fig:stab}.
Most of the relevant NS configurations (see Fig. \ref{fig:cool-h})
are then  affected by the presence of the quark core.
First we check the possibility of the 2SC+ normal quark phases.

\begin{figure}[ht]
\vspace{-0.5cm}
\centerline{
\psfig{figure=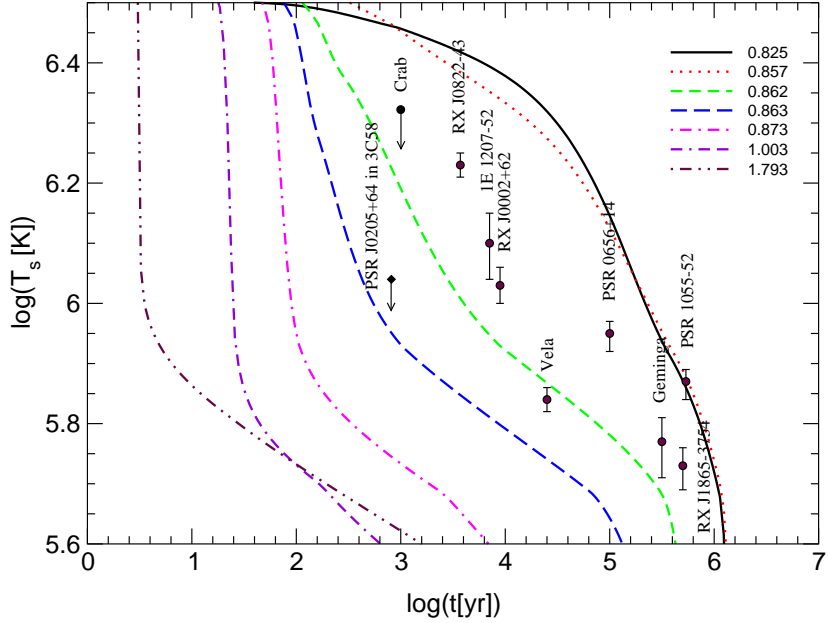,height=0.9\textwidth,angle=-90}}
\caption{Cooling curves for hybrid star configurations with Gaussian 
quark matter core in the 2SC phase.
The labels correspond to the gravitational masses of the configurations 
in units of the solar mass.}
\label{fig:cool-2sc}
\end{figure}

Fig. \ref{fig:cool-2sc} shows the cooling curves calculated with the 
Gaussian ansatz.  The variation of zero temperature gaps for the strong
pairing of quarks within 2SC phase in the interval $\Delta \sim
20\div 200~$MeV only slightly affects the results. The main
cooling process is the QDU process on normal quarks. We see that
the presence of normal quarks leads to too fast cooling. The data
could be explained only if all the masses lie in a very narrow
interval ($0.82<M/M_\odot<0.90$ in our case). 
The existence of only a very narrow mass interval in which the
data can be fitted seems us unrealistic as by itself as from the
point of view of the observation of the NS with different masses:
$M\simeq 1.41~M_{\odot}$ and $M\simeq 1.25~M_{\odot}$, cf.
\cite{L04}. Thus the data can't be satisfactorily explained.

\begin{figure}[ht]
\vspace{-0.5cm}
\centerline{
\psfig{figure=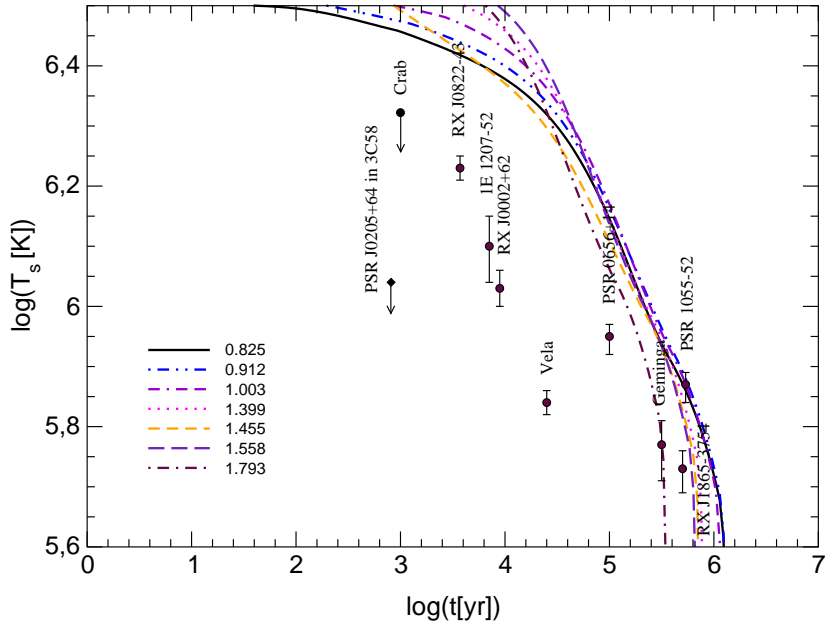,height=0.9\textwidth,angle=-90}}
\caption{Cooling curves for hybrid star configurations with Gaussian quark 
matter core in the 2SC+CSL phase. The weak pairing gap is 1 MeV.
The labels correspond to the gravitational masses of the configurations in 
units of the solar mass.}
\label{fig:cool-csl1}
\end{figure}

In Fig. \ref{fig:cool-csl1} we permit the weak pairing for all quarks which 
were assumed to be unpaired. We use $\Delta \simeq 1~$MeV for the
corresponding quark gap. Fig. \ref{fig:cool-csl1} demonstrates too slow 
cooling.

\begin{figure}[ht]
\vspace{-0.5cm}
\centerline{
\psfig{figure=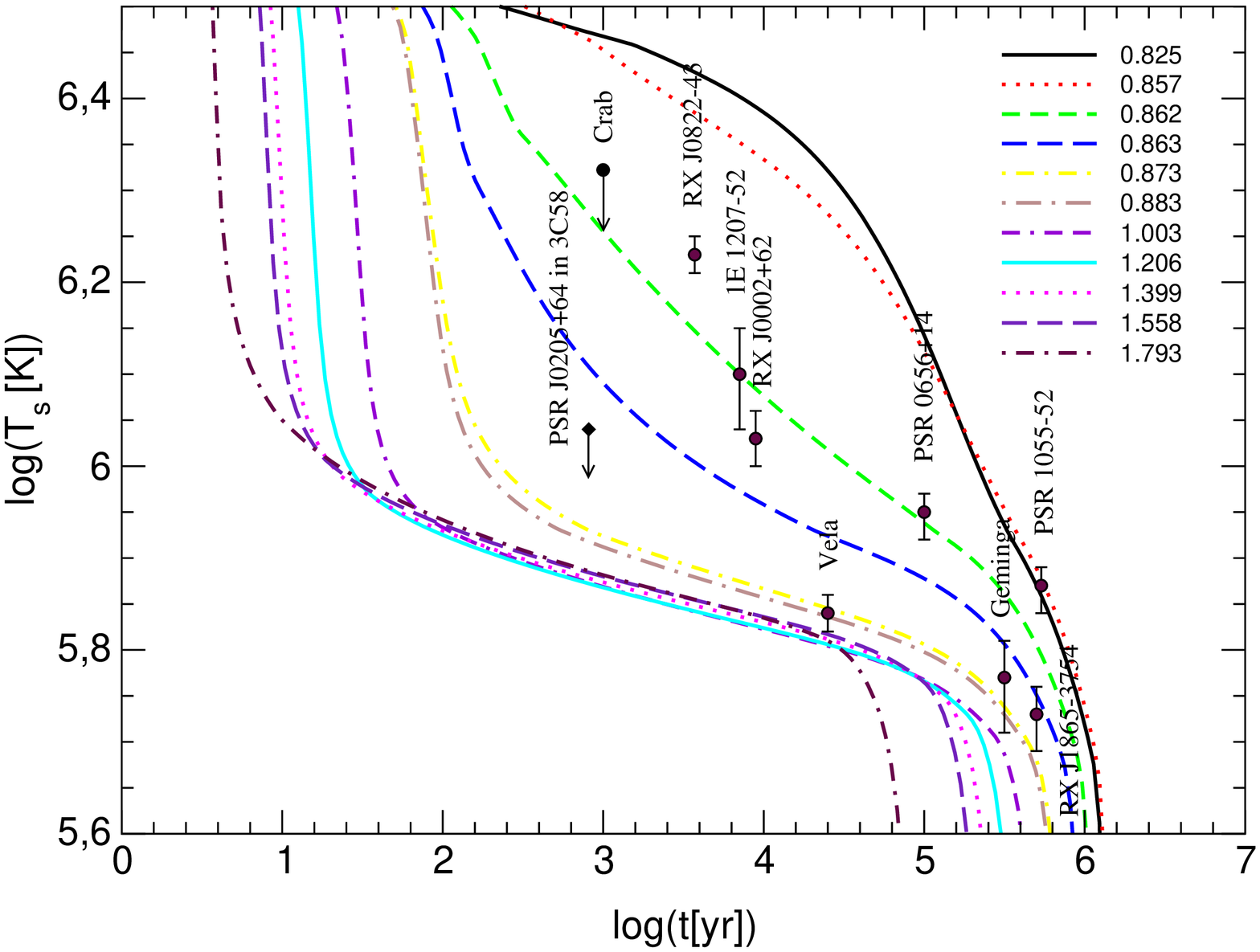,height=0.9\textwidth,angle=-90}}
\caption{Same as Fig. \ref{fig:cool-csl1} but 
with a weak pairing gap of 50 keV.}
\label{fig:cool-csl01}
\end{figure}

In Fig. \ref{fig:cool-csl01} we again allow for the weak pairing
of those quarks which were assumed to be unpaired in
Fig. \ref{fig:cool-2sc}, but now we use
$\Delta = 50~$ keV for the corresponding CSL quark gap. 
The data
are appropriately  fitted. The "slow cooling" data are explained
by the cooling of NS either without quark core ($M\sim0.5~M_\odot$,
see Fig. \ref{fig:cool-h}) or with a quark core.
Although the majority of experimental cooling points are covered by NS 
masses in a very narrow mass interval
$M=0.86\div 0.87 ~M_\odot$ the difference with Fig. \ref{fig:cool-2sc}
is crucial. We see that with the higher masses, up to $1.6~M_\odot$  
one may cover the rapid cooling point (Vela). Besides, the variation of 
parameters allows to shift all the curves up what permits to essentially 
broaden the mass interval that would cover the data. E.g. with the so called 
Tsuruta law for the $T_s-T_{in}$ relation the corresponding mass interval
is $M=0.86\div 1.79 ~M_\odot$. With these remarks we showed that the 
2SC+CSL+hadron scenario allows to fit the data as well as the purely 
hadronic scenario, see \cite{bgv2004}. 
To be more realistic one should further inlcude density dependences of the 
gaps what we intend to do in the forthcoming publication.   

\section*{Conclusion}
Concluding, we demonstrated that the present day cooling data can be
explained not only by a purely hadronic structure of NS interiors but 
also by a hybrid one with a complex  pairing pattern, where quarks
are partly strongly paired within the 2SC channel, and partly weakly paired 
within the CSL channel with gaps $\Delta \stackrel{<}{\sim} 50~$keV.
We conclude also that our choice of a density-independent weak pairing gap 
could be the reason why the mass interval for explaining slow and 
intermediate cooling data is very narrow. As it is well-known that the CSL gap
should have a strong density dependence \cite{BHO}, the fastening of the 
cooling by increasing the star masses should be partly compensated for
by the corresponding increase of the weak pairing gap. 
Corresponding calculations are under way.

\subsection*{Acknowledgement}

H.G. and D.V. acknowledge the hospitality and support of Rostock
University. The work of  H.G.  has been supported in part by the
Virtual Institute of the Helmholtz Association under grant No.
VH-VI-041, that of D.V. has been supported in part by DFG grant No.
436 RUS 17/117/03 and by RFBR grant NNIO-03-02-04008.
D.B. thanks the organizers of the workshop, in particular D.-K. Hong
and C.-H. Lee, for their invitation and support.


\end{document}